\documentclass[11pt, twoside, eqno]{article}
\usepackage{amssymb}
\usepackage{mathrsfs}
\usepackage{amsmath}
\usepackage{amsfonts}
\usepackage{amsthm}
\usepackage{subfigure}
\usepackage{graphicx}
\usepackage{booktabs}
\usepackage{setspace}

\allowdisplaybreaks \numberwithin{equation}{section}

\def\hs{\hspace{0.5cm}}

\newtheorem{obs}{\noindent\rm\bf Observation}

\begin{document}
%\doublespacing
\baselineskip=15pt
\renewcommand{\arraystretch}{2}
\arraycolsep=1pt
\title{\bf \Large  Multidimensional Zero-Correlation Linear Cryptanalysis of the Block Cipher KASUMI
\footnotetext{\hspace{-0.6cm} ${ }^{*}$ Corresponding authors. \\
   E-mail addresses: nlwt8988@gmail.com. }
\author{\vspace{-0.1cm}\bf Wentan Yi$^{*}$ and Shaozhen Chen  \\
\vspace{-0.5cm}\small\it State Key Laboratory of Mathematical Engineering and Advanced Computing,\\
\small\it Zhengzhou 450001, China }}
\date{}

\maketitle

\begin{center}
\begin{minipage}{15.2cm}
\small{\bf Abstract.}  The block cipher KASUMI, proposed by ETSI SAGE more than 10 years ago, is widely used for security in many synchronous wireless standards nowadays. For instance, the confidentiality and integrity of 3G mobile communications systems depend on the security of KASUMI.  Up to now, there are a great deal of cryptanalytic results on KASUMI, however, its security evaluation against the recent zero-correlation linear attacks is still lacking. In this paper, combining with some observations on the $FL$, $FO$ and $FI$ functions, we select some special input/output masks to refine the general 5-round zero-correlation linear approximations and propose the 6-round zero-correlation linear attack on KASUMI. Moreover, under the weak keys conditions that the second keys of the $FL$ function in round 2 and round 8 have the same value  at 1st to 8th and 11th to 16th bit-positions, we expand the attack to 7-round KASUMI(2-8). These weak keys take  1/$2^{14}$ of the key space.

\quad\, The new zero-correlation linear attack on the 6-round needs about $2^{85}$ encryptions with $2^{62.8}$ known plaintexts and $2^{54}$ memory bytes. For the attack under weak keys conditions on the last 7 round, the data complexity is about $2^{62.1}$ known plaintexts, the time complexity is about $2^{110.5}$  encryptions and the memory requirements are about $2^{85}$ bytes.
\medskip


\end{minipage}
\end{center}

\section{\large\bf Introduction}

With the rapid growth of wireless services, various security algorithms have been developed to provide users with effective and secure communications. The block cipher KASUMI, developed from a previous block cipher known as MISTY1[11], was chosen as the foundation for the 3GPP confidentiality and integrity  algorithm[15] and was also  recommended as the standard algorithms A5/3 and GEA3 in GSM and GPRS mobile communications systems[16]. For this reason, it is very important to understand the security offered by KASUMI.

Up to now, a great deal of attention have been paid to KASUMI and many cryptanalytic methods have been used to evaluate its security. In the single-key setting, Sugio et al.[12][14] gave the Integral-interpolation attack and higher order differential attack on 6/5-round KASUMI and later, they [13] improved the higher order differential attack on 5-round KASUMI to  practical complexity. K\"{u}hn [10] introduced the impossible differential attack on 6-round KASUMI.  At SAC 2012, Jia et al [9] refined the impossible differential by selecting some special input differential values and extended the 12-years old impossible differential attack on 6-round KASUMI to 7-round. In the related-key setting,
Blunden et al.[6] gave a related-key differential attack on 6-round KASUMI.
Using related-key booming and rectangle attack,
Biham et al. [5] proposed the first related-key attack on the full 8-round KASUMI. At Crypto 2010,  sandwich attacks [7], which belongs to a formal extension of booming attacks, was introduced to the full KASUMI.  As the key schedule of KASUMI is linear and simple, attack effects in the related-key setting are much better. However, these attacks assume control over the differences of two or more related keys, which renders the resulting attack inapplicable in most real world usage scenarios.
\begin{table}[tbp]
\centering
\scriptsize
\begin{tabular}{ccccccc}
\hline
Attack Type & Rounds & Date & Time & Memory & Source\\
\hline
\vspace{-0.1in}Higher-Order Differential & 5 & $2^{22.1}$CP & $2^{60.7}$Enc &$--$& [14] \\
\vspace{-0.1in}Higher-Order Differential&5 &$2^{28.9}$CP & $2^{31.2}$Enc &$--$& [13] \\
\vspace{-0.1in}Integral-Interpolation &6 &$2^{48}$CP & $2^{126.2}$Enc&$--$& [12]  \\
\vspace{-0.1in}Impossible Differential &6& $2^{55}$CP & $2^{100}$Enc &$--$& [10]  \\
\vspace{-0.1in}Multidimensional Zero-Correlation&6&$2^{62.8}$KP&$2^{85}$Enc &$2^{54}$ bytes& Sect.[4]\\
\vspace{-0.1in}Impossible Differential & 7(2-8) & $2^{52.5}$CP & $2^{114.3}$Enc &$--$ & [9]  \\
\vspace{-0.1in}Impossible Differential &7(1-7)& $2^{62}$KP & $2^{115.8}$Enc &$2^{84.4}$ bytes& [9]  \\
Multidimensional Zero-Correlation(WK)&7(2-8)&$2^{62.1}$KP&$2^{110.5}$Enc &$2^{85}$ bytes& Sect.[5]\\
\hline
\end{tabular}

CP,KP refer to the number of chosen plaintexts and known plaintexts.\\
Enc refers to the number of encryptions. $--$ means not given.\\
WK refers to the second keys of the $FL$ function in round 2 and 8 have the same value at 1-8 and 11-16 bit-positions.
\caption{Summary of the attacks on KASUMI}
\end{table}
In this paper, we  apply the  recent zero-correlation linear attacks  to the block cipher KASUMI.
Zero-correlation linear cryptanalysis, proposed by Bogdanov and Rijmen[1], is a novel promising attack technique for block ciphers.
It uses the linear approximation with correlation zero generally existing in block ciphers to distinguish between a
random permutation and a block cipher.  The initial distinguishers [1] had some limitations in terms of data complexity, which needs at least half of the codebook. In FSE 2012, Bogdanov and Wang [2] proposed a more data-efficient distinguisher by making use of  multiple linear approximations with correlation zero. The date complexity is reduced, however, the distinguishers rely on the assumption that all linear approximations with correlation zero are independent. To remove the unnecessary independency assumptions on the distinguishing side,  multidimensional distinguishers [3] had been constructed for the zero-correlation property at AsiaCrypt 2012.  Recently, the multidimensional zero-correlation linear cryptanalysis has been using in the analysis of the block cipher CAST-256[3], CLEFIA[4], HIGHT[17] and E2[18] successfully.

In this paper, we evaluate the security of KASUMI with respect to the multidimensional zero-correlation linear cryptanalysis. Our contributions can be summarized as follows:

1. The general 5-round zero-correlation linear approximations: $(\overline{\beta},0) \overset{\text{5-round}}{\longrightarrow} (\overline{\beta}, 0)$, that holds for any balanced Feistel scheme(the round function is bijective), can be used in the analysis of KASUMI, where $\overline{\beta} $ is any non-zero 32-bit value. However, if we take all non-zero values for $\overline{\beta}$, there will be two many guessed subkey bits involved in the key recovery process that the time complexity will be greater than exhaustive search. In order to reduce the number of guessed subkey bits, we only use some special zero-correlation linear approximations. We first investigate the properties of the linear masks propagate in  components (AND, OR functions)and then show some observations on the $FL$,$FO$ and $FI$ function. Based on those observations, we give some conditions the special linear approximations should satisfy.

2. We propose the multidimensional zero-correlation linear attack on 6-round  KASUMI. Up to now, there are no linear attacks on KASUMI and we bridge this gap, if we treat the zero-correlation linear attack as a special case of linear attacks.

3. We assume that the second keys of the $FL$ function in round 2 and round 8 have the same values at 1st to 8th and 11th to 16th bit-positions and then under this weak key conditions, we expand the attack to 7 rounds(2-8).

The paper is organized as follows: we list some notations,  give a brief description of the block cipher KASUMI  and outline the ideas of the multidimensional zero-correlation linear cryptanalysis in Section 2. Some observations on AND, OR, $FL$ $FO$ and $FI$  functions are shown in Section 3. Section 4 and Section 5 illustrate our attacks on 6-round and 7-round KASUMI. We conclude this paper in Section 6.

\section{\large \bf  Preliminaries}
\subsection{\bf Notations}

%Throughout this paper, we will use some symbols, which are listed as follows:

\quad \ $FL_i$ \quad \quad :  the $i$-th $FL$ function of KASUMI with subkey $KL_i$.

$FO_i$ \quad \quad  :  the $i$-th $FO$ function of KASUMI with subkey $(KO_i, KI_i)$.

$FI_{ij}$ \quad \quad : the $j$-th $FI$ function of $FO_i$ with subkey $KI_{i j}$.

 $x\ggg i$ \quad : $x$ rotates left by $i$ bits.

 $x\lll i$ \quad : $x$ rotates right by $i$ bits.

$\wedge$ \quad \quad\quad\,: bitwise AND.

$\vee$ \quad\quad \quad\,: bitwise OR.

$\oplus$\quad\quad \quad\ : bitwise XOR.

$\neg$ \quad\quad \quad\,: bitwise NOT.

$a\cdot b$ \quad\quad \,: the scalar product of binary vectors by $a\cdot b = \oplus_{i=0}^{n-1}a_i b_i$.

$a\diamond b$ \quad\ \ \, \, \,: the bitwise point multiplication of binary vectors by $a\diamond b =(a_0 b_0, a_1 b_1,..., a_{n-1} b_{n-1})$.

$X\|Y$ \ \,\quad: the concatenation of $X$ and $Y$.

$z[i]$ \quad \ \quad: the $i$-th bit of $z$, and $'0'$ is the most significant bit.

$z[i_1-i_2]$ : the $(i_2 - i_1 + 1)$ bits from the $i_1$-th bit to $i_2$-th bit of $z$.

\subsection{\bf Description of KASUMI}
The KASUMI algorithms [15] are symmetric block ciphers with a block size of 64 bits and a key size of 128 bits. We give a brief description of KASUMI in this section.
\begin{figure}
  \centering
  \includegraphics[width=12cm]{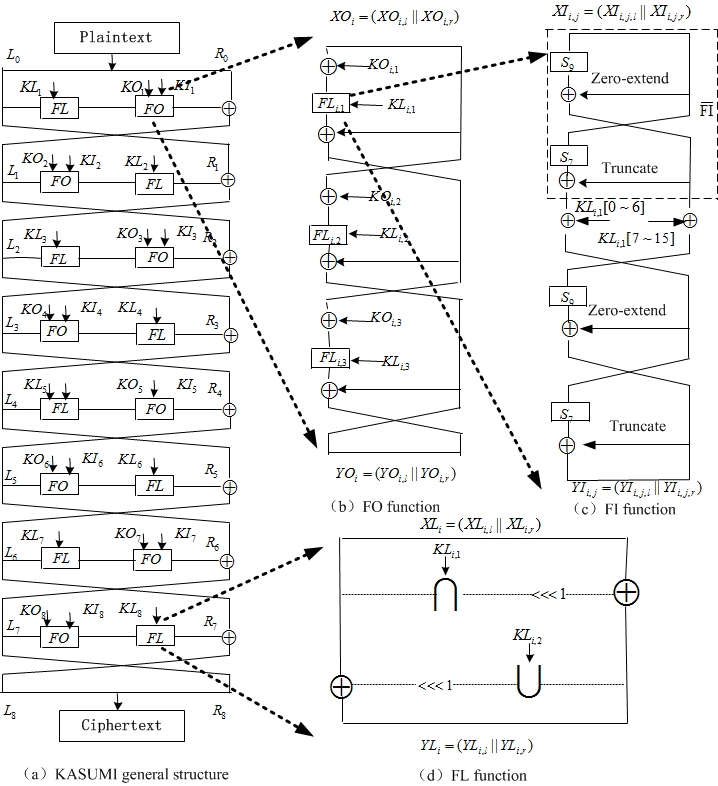}
  \caption{The structure and building blocks of KASUMI}
\end{figure}

KASUMI is a Feistel structure with 8 rounds, see Fig. 1 (a) for an illustration. The round function consists of an $FL$ function and an $FO$ function. The $FL$ function is a simple key-dependent boolean function, depicted in
Fig. 1 (d). Let the inputs of the $FL$ function of the $i$-th round be $XL_i =
XL_{i,l}\|XL_{i,r}, KL_i =(KL_{i,1}, KL_{i,2})$, the output be $YL_i = YL_{i,l}\|YL_{i,r}$, where
$XL_{i,l}$,$XL_{i,r}$,$YL_{i,l}$ and $YL_{i,r}$ are 16-bit integers. We define the $FL$ function as follows:
$$YL_{i,r} =((XL_{i,l} \wedge KL_{i,1})\lll1) \oplus XL_{i,r};$$
$$YL_{i,l} =((YL_{i,r} \vee KL_{i,2})\lll1) \oplus XL_{i,l}, $$

\begin{table}[tbp]
\centering
\scriptsize
\begin{tabular}{l}
\hline
Algorithm 1 The KASUMI block cipher \\
\hline
\vspace{-0.1in}Require: 64-bit plaintext $P=(L_0, R_0)$; main key $K$,\\
\vspace{-0.1in}Ensure: 64-bit ciphertext $C=(L_8,R_8)$.\\
\vspace{-0.1in}1: Derive  round keys $KO_i$, $KI_i$ and $KL_i$ $(1 \leq i \leq 8)$ from $K$.\\
\vspace{-0.1in}2: for $j=1$ to 8 do\\
\vspace{-0.1in}3: if $j$ is odd, do \\
\vspace{-0.1in}4: $L_j = FO(FL(L_{j-1},KL_j),KO_j,KI_j)\oplus R_{j-1}, R_j = L_{j-1},$ \\
\vspace{-0.1in}5: else, do :\\
\vspace{-0.1in}6: $L_j = FL(FO(L_{j-1},KO_j,KI_j),KL_j)\oplus R_{j-1}, \, R_j = L_{j-1}.$\\
\vspace{-0.1in}7: end for \\
8: return $C=(L_8, R_8)$.\\
\hline
\end{tabular}
\end{table}
The $FO$ function, depicted in Fig. 1 (b), is another three-round Feistel structure consisting of three $FI$ functions and key
mixing stages. Let $XO_i = XO_{i,l}\|XO_{i,r}$, $KO_i =
(KO_{i,1},KO_{i,2},KO_{i,3})$, $KI_i =(KI_{i,1}, KI_{i,2}, KI_{i,3})$ be the inputs of the $FO$
function of $i$-th round, and $YO_i = YO_{i,l}\|YO_{i,r}$ be the corresponding output,
where $XO_{i,l}$,$XO_{i,r}$,$YO_{i,l}$,$YO_{i,r}$ and $\overline{XI_{i,3}}$ are 16-bit integers. Then the $FO$
function has the form
$$\overline{XI_{i,3}} = FI((XO_{i,l}\oplus KO_{i,1}),KI_{i,1})\oplus XO_{i,r};$$
$$YO_{i,l} = FI((XO_{i,r} \oplus KO_{i,2}),KI_{i,2}) \oplus \overline{XI_{i,3}};$$
$$YO_{i,r} = FI((\overline{XI_{i,3}} \oplus KO_{i,3}),KI_{i,3})\oplus YO_{i,l}.$$

The $FI$ function uses two S-boxes $S_7$ and $S_9$ which are permutations of 7-bit to
7-bit and 9-bit to 9-bit respectively. Suppose the inputs of the $j$-th $FI$ function
of the $i$-th round are $XI_{i,j}$, and the output is $YI_{i,j}$, where $XI_{i,j}$ and $YI_{i,j}$
are 16-bit integers. We define half of $FI$
function as $\overline{FI}$, which is a 16-bit to 16-bit permutation. The structure of $\overline{FI}$ and
$FI$ is depicted in Fig. 1 (c).
$\overline{YI_{i,j}} = FI(XI_{i,j})$ is defined as
$$\overline{YI_{i,j}} [7 - 15] = S_9(XI_{i,j}[0 - 8]) \oplus XI_{i,j} [9 -15];$$
$$\overline{YI_{i,j} }[0- 6] = S_7(XI_{i,j}[9-15]) \oplus \overline{YI_{i,j} }[7 - 15],$$
the $FI$ function is simplified as
$$YI_{i,j} = FI(XI_{i,j} ,KI_{i,j} )= \overline{FI}((\overline{FI}(XI_{i,j} ) \oplus KI_{i,j} )\lll7).$$

Let $L_i||R_i=\big((L_{i,l}\|L_{{i},r})\|(R_{i,l}\|R_{{i},r})\big)$ be the input of the $i$-th round, and then the round function is defined as
$$L_i = FO(FL(L_{i-1},KL_i),KO_i,KI_i)\oplus R_{i-1}, R_i = L_{i-1},$$
where $i =1, 3, 5, 7$, and when $i =2, 4, 6,8,$
$$L_i = FL(FO(L_{i-1},KO_i,KI_i),KL_i)\oplus R_{i-1}, \, R_i = L_{i-1}.$$
Here, $(L_0,R_0)$, $(L_8,R_8)$ are the plaintext and ciphertext respectively, and $L_{i-1}$,
$R_{i-1}$ denote the left and right 32-bit halves of the $i$-th round input.  The KASUMI cipher can be described in Algorithm 1.

The key schedule of KASUMI is much simpler than the original
key schedule of MISTY1. The 128-bit key $K$ is divided into eight 16-bit words:
$(k_1,k_2, ..., k_8)$, i.e., $K =(k_1,k_2,k_3,k_4,k_5,$ $k_6,k_7,k_8)$. In each round, eight key
words are used to compute the round subkeys, that is,  $KL_i$, $KO_i$ and $KI_i$, where $KL_i =(KL_{i,1},KL_{i,2}), KO_i=(KO_{i,1},KO_{i,2},KO_{i,3})$ and $KI_i =(KI_{i,1}, KI_{i,2}, KI_{i,3})$. We summarize the details of the key schedule of KASUMI in Tab. 2.

\subsection{\bf Zero-correlation Linear cryptanalysis}

In this section, we briefly recall the basic concepts of multidimensional zero-correlation linear cryptanalysis.
  Consider a function $f : F^n_2 \mapsto F^m_2$ and let the input of the function be $x\in F_2^n$. A linear approximation
with an input mask $\alpha$ and an output mask $\beta$ is the following function:
$$x \mapsto \beta \cdot f(x)\oplus a\cdot x,$$
and its correlation  is defined as follows
$$C(\beta \cdot f(x), a\cdot x)=2Pr_{x}(\beta \cdot f(x)\oplus a\cdot x=0)-1.$$

\begin{table}[tbp]
\centering
\scriptsize
\begin{tabular}{ccccccccc}
\hline
Round & $KL_{i,1}$ & $KL_{i,2}$ & $KO_{i,1}$ & $KO_{i,2}$ & $KO_{i,3}$ & $KI_{i,1}$ & $KI_{i,2}$ & $KI_{i,3}$ \\
\hline
\vspace{-0.1in}1&$k_1\lll1$ &$k'_3$ &$k_2\lll5$ &$k_6\lll8$ &$k_7\lll13$ &$k'_5$&$k'_4$& $k'_8$\\
\vspace{-0.1in}2&$k_2\lll 1$ &$k'_4$ &$k_3\lll5$ &$k_7\lll8$ &$k_8\lll13$ &$k'_6$&$k'_5$& $k'_1$\\
\vspace{-0.1in}3&$k_3\lll 1$ &$k'_5$ &$k_4\lll5$ &$k_8\lll8$ &$k_1\lll13$ &$k'_7$&$k'_6$& $k'_2$\\
\vspace{-0.1in}4&$k_4\lll 1$ &$k'_6$ &$k_5\lll5$ &$k_1\lll8$ &$k_2\lll13$ &$k'_8$&$k'_7$& $k'_3$\\
\vspace{-0.1in}5&$k_5\lll 1$ &$k'_7$ &$k_6\lll5$ &$k_2\lll8$ &$k_3\lll13$ &$k'_1$&$k'_8$& $k'_4$\\
\vspace{-0.1in}6&$k_6\lll 1$ &$k'_8$ &$k_7\lll5$ &$k_3\lll8$ &$k_4\lll13$ &$k'_2$&$k'_1$& $k'_5$\\
\vspace{-0.1in}7&$k_7\lll 1$ &$k'_1$ &$k_8\lll5$ &$k_4\lll8$ &$k_5\lll13$ &$k'_3$&$k'_2$& $k'_6$\\
8&$k_8\lll 1$ &$k'_2$ &$k_1\lll 5$ &$k_5\lll8$ &$k_6\lll13$ &$k'_4$&$k'_3$& $k'_7$\\
\hline
\end{tabular}

$x\lll i$: $x$ rotates left by $i$ bits.
$k'_i=k_i\oplus c_i$, where the $c_i$s are fixed constants.
\caption{The key schedule of KASUMI}
\end{table}

In zero-correlation linear cryptanalysis, the distinguishers use linear approximations with zero correlation for
all keys while the classical linear cryptanalysis utilizes linear approximations with correlation far from zero. Bogdanov et al. [3] proposed a multidimensional zero-correlation linear distinguisher using $\ell$ zero-correlation linear approximations and requiring $O(2^n/\sqrt{\ell})$ known plaintexts, where $n$ is the block size of a cipher.

We treat the zero-correlation linear approximations available as a linear
space spanned by $m$ base zero-correlation linear approximations such that all
$\ell=2^m $ non-zero linear combinations of them have zero correlation. For
each of the $2^m$ data values $z \in F_2^m $, the attacker initializes a counter $V[z]$, $z=0, 1,...,2^m-1$ to value zero.
Then, for each distinct plaintext, the attacker computes the data value $z$ in $F^m_2$ by evaluating the $m$ basis linear
approximations, that is, $z[i]=\alpha_i\cdot p \oplus \beta_i \cdot c$ , $i=0,...,m-1$, where we denote the $i$-th basis linear
approximation and any plaintext-ciphertext pair by $(\alpha_i,\beta_i)$ and $(p, c)$. Then, increase the counter $V[z]$ of this data value by one. Then the attacker computes the statistic $T$:
$$T=\sum_{i=0}^{2^m-1}\frac{(V[z]-N2^{-m})^2}{N2^{-m}(1-2^{-m})}.\eqno(1)$$
The statistic $T$  follows a $\mathscr{X}^2$ -distribution with mean $\mu_0=(\ell-1)\frac{2^n-N}{2^n-1}$ and variance
$\sigma^2_0=2(\ell-1)\big(\frac{2^n-N}{2^n-1}\big)^2$ for the right key guess, while for the wrong key guess,
it follows a $\mathscr{X}^2$-distribution with mean $\mu_1=\ell-1$ and variance $\sigma_1^2=2(\ell-1)$.

If we denote the probability of false positives and the probability of false negatives to distinguish between a wrong key and a right key as $\beta_0$ and $\beta_1$, respectively, and we consider the decision threshold $\tau =\mu_0+\sigma_0z_{1-\beta_0}=\mu_1-\sigma_{1}{z_{1-\beta_1}}$, then the number of known plaintexts $N$ should be about
$$N=\frac{(2^n-1)(z_{1-\beta_0}+z_{1-\beta_1})}{\sqrt{(\ell-1)/2}+z_{1-\beta_0}}+1,\eqno(2)$$
where $z_{1-\beta_0}$ and $z_{1-\beta_1}$are the respective quantiles of the standard normal distribution, see [3] for detail.

\subsection{\bf  The  Partical-Sum  technique}
The partial-sum technique [8] was first introduced by Ferguson et al. to analyse the block cipher AES.
The partial-sum technique can reduce the complexity by  partially computing the sum by guessing each key one after another.
For an example, in the key recovery phase of the AES, the partial decryption involves 4 bytes of the key and 3 bytes of the ciphertext.  We denoted the byte position $i$ of each ciphertext and corresponding keys by $c_{i}$, $k_i$ and suppose that
$2^{24}$ ciphertexts to be analyzed, then, the equation can be described as follows:
$$\bigoplus _{n=1}^{2^{24}}\Big[S_{3}\big(S_{2}(c_{2,n}\oplus k_2)\oplus S_1(c_{1,n}\oplus k_1)\oplus S_0(c_{0,n}\oplus k_0\big)\oplus k_3)\Big].\eqno(3)$$

With a straightforward method, the analysis takes $2^{24+32}=2^{56}$ partial decryptions, while the partial-sum technique requires about $2^{41.6}$ partial decryptions. The idea is partially computing the sum by guessing each key byte one after another.

1. Guess two key bytes $k_2$ and $k_1$. Allocate a counter $N_1[x_1]$ for each of $2^{16}$ possible values of
$x_1=x_1^0\|x_1^1$ and set them zero. For $2^{24}$ ciphertexts $(c_{0,n},c_{1,n},c_{2,n})$, compute $x_1^1=S_2(c_{2,n}\oplus k_2) \oplus S_1(c_{1,n} \oplus k_1)$ and let $x_1^0=c_{0,n}$, calculate the number of ciphertext with given values $x_1$ and save it in $N_1[x_1]$.

2. Guess the key byte $k_0$. Allocate a counter $N_2[x_2]$ for each of $2^{8}$ possible values of
$x_2$, and set them zero. For $2^{8}$ possible values $c_{0,n}$, compute $x_2=x_1^1\oplus S_0(c_{0,n}\oplus k_0)$, and update the value $N_2[x_2] =N_2[x_2]+ N_1[x_1]$.

3. Guess the key byte $k_3$. Allocate a counter $N_3[x_3]$ for each of $2^{8}$ possible values of
$x_3$ and set them zero. For $2^{8}$ possible values $x_2$, compute $x_3= S_3(x_2\oplus k_3)$, and update the value $N_3[x_3] =N_3[x_3]+ N_2[x_2]$.

In the first step, $k_2$ and $k_1$ are guessed, the complexity is $2^{16}\times 2^{24}=2^{40}$. For the second step, $k_0$ is guessed, the complexity is $2^{16}\times 2^{8}\times 2^{16}=2^{40}$. Finally, $k_3$ is guessed and Eq. (3) is computed. The complexity for the guess of
$k_3$ is $2^{16} \times  2^{8}\times 2^{8}\times 2^{8} =2^{40}$.

\section{\large\bf Some Observations in KASUMI }
\begin{figure}
\centering
\includegraphics[width=12cm]{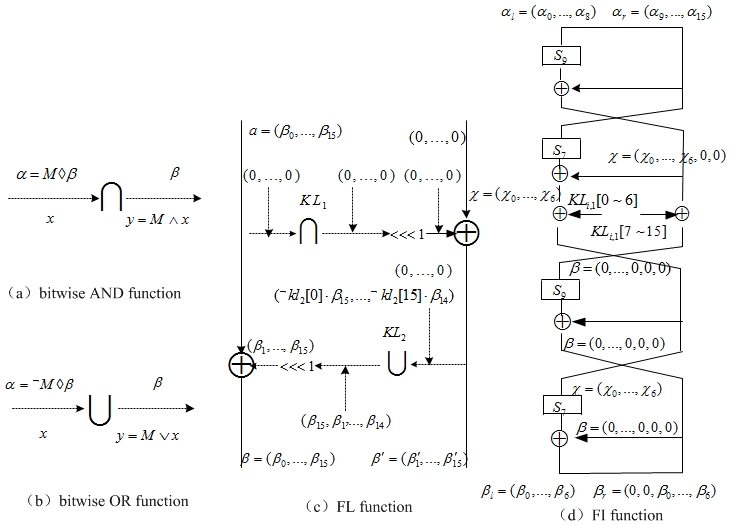}
\caption{Property of  AND, OR, $FL$ and $FI$ functions}
\end{figure}

We describe some observations on AND, OR,  $FL$, $FO$ as well as $FI$ functions, which will be used in our cryptanalysis of KASUMI.
\begin{obs}
Let $M$ be a $\ell$-bit value and define the OR, AND functions $h_1$, $h_2$ as $h_1(x)=M \vee x$, $h_2(x)=M \wedge x$. Then there are two properties of the two functions, such that

\begin{itemize}
\item[\rm (I)]  For any $\ell$-bit masks $\alpha$ and $\beta$, $C(\beta \cdot h_1(x),\alpha\cdot x)\neq 0$ if and only if $ \alpha= \urcorner M\diamond \beta$;
\item[\rm (II)] For any $\ell$-bit masks $\alpha$ and $\beta$, $C(\beta\cdot h_2(x),\alpha\cdot x)\neq 0$ if and only if $ \alpha= M\diamond \beta$.
\end{itemize}
\end{obs}

\proof  We only consider the case of the $OR$ function $h_1(x)$. Noticed that the $j$-th bit of the output mask $\beta[j]$  does not affect the $i$-th bit of the input mask $\alpha[i]$ when $i\neq j$, then, for any $i\in (0,\, \ell-1)$, we have that
$\alpha[i]=\beta[i]$, when $M[i]=0$, and $\alpha[i]=0$, when $M[i]=1$, which can be summarized as $\alpha= \urcorner M\diamond \beta$. The necessary property can be  established clearly. See Figure 2 (b).

\begin{obs}
If the output masks of $FL_i$ function is $(\beta, \beta' )$, where $\beta'=(\beta\ggg 1)\diamond KL_{i, 2} $, that is,

$$\beta'[j]=\urcorner KL_{i,2}[j] \beta[j+1],\  \text{where}\ j=0,2,...14, \text{and} \ \beta'[15]=\urcorner KL_{i,2}[15]\beta[0],$$
then the input masks of $FL_i$ function is $(\beta, 0)$, see Figure 2(c).
\end{obs}

\begin{obs}
If the output masks of $FI_{i,j}$ function is $\beta$, where $\beta$ is a $16$-bit value with $\beta[0-6]=\beta[9-15]$ and $\beta[7]= \beta[8]=0$, then the input masks of $FI_{i,j}$ function $\alpha$ only depend on 7-bit subkey $KI_{i,j}[0-6]$, see Figure 2(d).
\end{obs}

Base on Observation 2, Observation 3 and the structure of round functiom of the KASUMI block cipher, we have the following two results.

\begin{obs}
 Let $ (\beta, \beta')$ be  the output mask of $FL_6$ function, such that $\beta'=(\beta\ggg 1)\diamond \urcorner KL_{6,2} $, then the input mask of $FO_{6,1}$  function $(\alpha, \alpha')$ only depend on the 64-bit subkey $KO_{6,1}$, $KI_{6,1}$, $KO_{6,2}$, and $KI_{6,2}$. Moreover, if the equations $\beta[0-6]=\beta[9-15]$ and $\beta[7]=\beta[8]=0$ hold, then $(\alpha, \alpha')$ only depend on the 46-bit subkey $KO_{6,1}$, $KI_{6,1}[0-6]$, $KO_{6,2}$, and $KI_{6,2}[0-6]$.
\end{obs}

\begin{obs}
Let $(\beta, \beta' )$ be the output mask of $FL_2$ and $FL_8$ functions, such that $ \beta'=(\beta\ggg 1)\diamond \urcorner KL_{2,2}\diamond \urcorner KL_{8,2}$, and for any $0\leq i\leq 15$,

\begin{displaymath}
(\beta\ggg 1)[i] = \left\{ \begin{array}{ll}
0,\quad & \textrm{if $(KL_{2,2}\oplus KL_{8,2})[i]=1$};\\
0 \ \text{or}\  1, \quad & \textrm{if $(KL_{2,2}\oplus KL_{8,2})[i]=0$}.
\end{array} \right.
\end{displaymath}
then the output mask of $FI_{2,3}$ and $FI_{8,3}$ be zero, and the input masks of $FO_{2}$, $FO_{8}$ functions $(a, a')$, $(\gamma, \gamma')$ depend on the 96-bit subkeys $KO_{2,1}$, $KI_{2,1}$, $KO_{2,2}$, $KI_{2,2}$, $KO_{8,1}$, $KI_{8,1}$, as $KI_{8,2}$,$KO_{8,2}$ can be deduced from above subkeys.  Moreover, if the equations $\beta[0-6]=\beta[9-15]$ and $\beta[7]=\beta[8]=0$ hold, then $(\alpha, \alpha')$ only depend on the 78-bit subkeys $KO_{2,1}$, $KI_{2,1}[0-6]$, $KO_{2,2}$, $KI_{2,2}$, $KO_{8,1}$, $KI_{8,1}[0-6]$.
\end{obs}
%\begin{equation*}
%\begin{split}
%(\beta\ggg 1)[i]=0, \text{when}\  , \quad \quad i=0,1,2,...15;\\
%(\beta\ggg 1)[i]=0 \ \text{or}\  1, \text{when}\  , \quad i=0,1,2,...15,
%\end{split}
%\end{equation*}

%\begin{figure}
%  \centering
%  \includegraphics[width=4cm]{fl.png}
%  \caption{the one picture}
%\end{figure}

\section{\large\bf Key-Recovery Attack on 6-Round  KASUMI}
The generic 5-round zero-correlation linear approximations of the balanced Feistel structure was introduced by Bogdanov and Rijmen in [1], which is: $(\overline{\beta}, 0)\overset{\text{5-Round}}{\longrightarrow}(\overline{\beta}, 0)$, where $\overline{\beta}$ is a 32-bit non-zero value. Combined with the  Feistel structure of the round function, some special values of  input/output masks $\beta$ are selected to attack the 6-round version of KASUMI. We mount the 5-round zero-correlation linear approximations from round 1 to round 5, and extend one round  backward. We select the 5-round zero-correlation linear approximations as:
$$(\beta\| \beta', 0)\overset{\text{5-Round}}{\longrightarrow}(\beta\| \beta', 0),$$
where  $\beta'=(\beta\ggg 1)\diamond \urcorner KL_{6,2}$ and  $\beta$ is 16-bit non-zero value with $\beta[0-6]=\beta[9-15]$ and $\beta[7]=\beta[8]=0$.
\begin{figure}
  \centering
  \includegraphics[width=12cm]{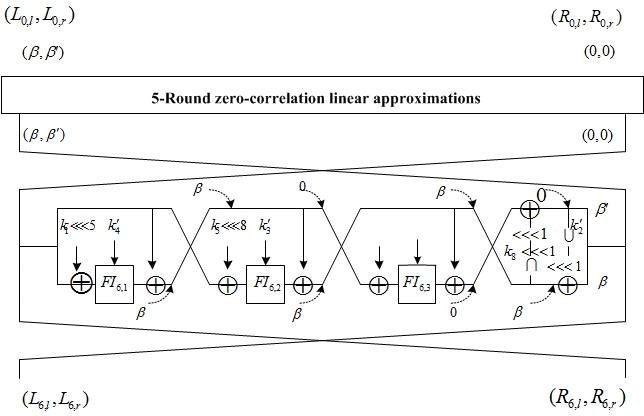}
  \caption{Multidimensional Zero-correlation attack on 6-round KASUMI}
\end{figure}

The choice  is to minimize the
key words guessing during the attack on 6-round  KASUMI.
Based on observations 3, we know that, if the input mask of the
first round is selected as above, $KI_{6,3}$, $KO_{6,3}$ and parts of $KI_{6,1}$ and $KI_{6,2}$ are not involved in the computation, which can help us to reduce the complexity of the attack.
The zero-correlation linear attack on  6-round  KASUMI is demonstrated as follows, see also Fig. 3.

%In this section, we describe our attacks on 6 rounds of KASUMI. We use the FFT to reduce time complex in our attack. The number of guessed key bits is affected by several parameters including the zero-correlation linear property we choose (values of ¦Á and ¦Â), the position of the property(rounds spanned by zero-correlation approximations), and
%
%the number of rounds added before and after this property. To optimize the attack complexities, a proper choice
%of these parameters is needed. We have implemented the
%search for the best parameters in a computer program
%which counts the number of guessed key bits in the par-
%tial encryption/decryption phase for all possible combina-
%tions of the parameters. To reduce the time complexity, we
%choose parameters with the least number of guessed key
%bits.

In our attack, we guess the subkeys and evaluate the linear approximation $(\beta,\beta')^{T}\cdot \big((L_{0,l}\oplus L_{6,l}), (L_{0,r} \oplus L_{6,r})\big)=0$, that is
$$(\beta,\beta')\cdot(L_{0,l}\oplus L_{6,l}\oplus R_{6,r},L_{0,r}\oplus L_{6,r})\oplus \beta\cdot\big(FI(R_{6,l}\oplus( k_1\lll 5), k'_4) \oplus FI(R_{6,r}\oplus (k_5\lll8), k'_3)\big)=0,$$
where  $\beta'=(\beta\ggg 1)\diamond \urcorner KL_{6,2}$ and  $\beta$ is any 16-bit non-zero value with $\beta[0-6]=\beta[9-15]$ and $\beta[7]=\beta[8]=0$.
 Then the key recovery attack on 6-round KASUMI is proceeded with Partial-sum technique as follows:

1. Collect $N$ plaintexts with corresponding ciphertexs.  Allocate a 16-bit counter $N_0[x_0]$ for each of $2^{53}$ possible values of $x_0=x_0^1\|x_0^2\|x_0^3\|x_0^4$, where $x_0^1 =R_{6,l}$, $x_0^2=R_{6,r}$, $x_0^3=(L_{0,r}\oplus L_{6,r})[0-7]\|(L_{0,r}\oplus L_{6,r})[10-15]$, $x_0^4=(L_{0,l}\oplus L_{6,l}\oplus R_{6,r})[0-6]\oplus (L_{0,l}\oplus L_{6,l}\oplus R_{6,r})[9-15]$ and set them zero. Calculate the
number of pairs of plaintext-ciphertext with given values $x_0$ and save it in $N_0[x_0]$. In
this step, around $2^{64}$ plaintext-ciphertext pairs are divided into $2^{53}$ different states. So the assumption $N_0$ as a 16-bit counter is sufficient.

2. Guess the 14-bit $KL_{6,2}[0-7], KL_{6,2}[10-15]$. Allocate a counter $N_1[x_1]$ for each of $2^{39}$ possible values of
$x_1=x_1^1\|x_1^2\|x_1^3$, where $x_1^1=x_0^1 $, $x_1^2=x_0^2$ and set them zero. For $2^{14} $ possible values of $x_{0}^3$, compute $x_1^3=x^4_0\oplus\big((\neg (KL_{6,2}[0-7]\|KL_{6,2}[10-15])\diamond x_0^3)\lll1\big)[0-6]\oplus\big((\neg (KL_{6,2}[0-7]\|KL_{6,2}[10-15])\diamond x_0^3)\lll1\big)[9-15] $ and update the value $N_1[x_1] = N_1[x_1] + N_0[x_0]$.

3. Guess the 23-bit $KO_{6,1}$ and $KI_{6,1}[0-6]$. Allocate a counter $N_2[x_2]$ for each of $2^{23}$ possible values of
$x_2=x_2^1\|x_2^2$, where $x_2^1 = x_1^2$ and set them zero. For all $2^{16} $ possible values of $x_1^1$, compute $x_2^2=x_1^3\oplus FI_{6,1}(x_1^1\oplus KO_{6,1}, KI_{6,1})[0-6]\oplus FI_{6,1}(x_1^1\oplus KO_{6,1}, KI_{6,1})[9-15]$ and update the value $N_2[x_2] = N_2[x_2] + N_1[x_1]$.

4. Guess the 23-bit $KO_{6,2}$ and $KI_{6,2}[0-6]$. Allocate a counter $N_3[x_3]$ for each of $2^{7}$ possible values of
$x_3$  and set them zero. For all $2^{16} $ possible values of $x_{2}^1$, compute $x_3^1=x_2^2\oplus FI_{6,2}(x_2^1\oplus KO_{6,2}, KI_{6,2})[0-6]\oplus FI_{6,2}(x_2^1\oplus KO_{6,2}, KI_{6,2})[9-15]$ and update the value $N_3[x_3] = N_3[x_3] + N_2[x_2]$.

5.  Allocate a 64-bit counter vector $V[z]$ for 7-bit $z$, where $z$ is the concatenation of evaluations of 7 basis zero-correlation masks.
Compute $z$ from $x_3$ with 7 basis zero-correlation masks, save it in $N[z]$, that is $N [z]+ = N_3[x_3]$. Compute the statistic $T$ according to Equation (1). If $T<\tau$ , the guessed key value is a right key candidate.

6. As there are 68 master key bits that we have not guessed, we do exhaustive search for all keys conforming to this possible key candidate.
%\begin{table}[tbp]
%\centering
%\small
%\begin{tabular}{cccccc}
%\hline
%Step & Guess & Computed States & Counter-Size \\
%\hline
%\vspace{-0.1in}1&$k_2$ &$x^1_i=(L_{5,l}|L_{5,r}|X^1)$ &48  \\
%\vspace{-0.1in}2&$k_4$,$k_1$ &$x^2_i=(L_{5,r}|X^2)$ &32  \\
%3&$k_5$,$k_3$ &$x_i^3=(X^3)$ &16  \\
%\hline
%\end{tabular}
%\caption{Partial decryption on 6-Round KASUMI}
%\end{table}

In this attack, we set the type-I error probability $\beta_0 =2^{-2.7}$ and the type-II error probability $\beta_1 =2^{-45}$. We have
$z_{1-\beta_0}\approx 1$, $z_{1-\beta_1}\approx 3.3$, $n = 64$, $ l=2^{7}$. The date complex $N$ is about $2^{62.8}$ and the decision threshold $\tau \approx 2^{6.22}$.

During the encryption and decryption phase, there are about $2^{60}\times 2^{-45}$ = $2^{15}$ key candidates survive in the wrong key filtration. The complexity of Step 2,3,4 is no more than $2^{14+53}=2^{67}$ memory access, $2^{14+23+39}=2^{76}$ memory access and $2^{14+23+23+23}={83}$ memory access, respectively. If we consider one memory accesses as a 6-round encryption, the total time complexity of step 3,4,5 is about $2^{83}$ of 6-round KASUMI. The complexity of Step 6 is about $2^{85}$ 6-round KASUMI encryptions which is also the dominant part of our attack. In total, the data complexity is about $2^{62.1}$ known plaintexts, the time complexity is about $2^{85}$ of 6-round encryptions and the memory requirements are $2^{54}$ bytes for counters.

\section{\bf Key-Recovery Attack on  7-Round KASUMI(2-8)}

In this section, we extend our attacks to the last 7 round of KASUMI. We mount the 5-round zero-correlation linear approximations from round 3 to round 7, and extend one round forward and backward respectively. We assume that the subkeys $KL_{2,2}[0-7]=KL_{8,2}[0-7]$ and $KL_{2,2}[10-15]=KL_{8,2}[10-15]$. There are $2^{-14}$ of the master keys space having this property. In the attack, based on Observation 5, we also select some special input/output masks to reduce number of guessed key bits. In our attack, we guess the subkeys and evaluate the linear approximation $(\beta,\beta')^{T}\cdot \big((L_{2,l},L_{2,r})$ $ \oplus (R_{7,l},R_{7,r})\big)=0$, that is
\begin{eqnarray*}
(\beta,\beta')\cdot \big((R_{1,l}\oplus L_{8,l}\oplus L_{1,r}\oplus R_{8,r}), (R_{1,r}\oplus L_{8,r})\big)\oplus
 \beta\cdot \big(FI(L_{1,l}\oplus (k_3\lll 5), k'_6) \oplus FI(L_{1,r} \\  \oplus k'_5,  k_7\lll 8)
\oplus FI(R_{8,l}\oplus (k_1\lll 5), k'_4) \oplus FI(R_{8,r}\oplus (k_5\lll 8), k'_3)\big)=0,
\end{eqnarray*}
where $ \beta'=\beta\ggg 1\diamond \urcorner KL_{8,2}$ and  $\beta$ is 16-bit non-zero value with $\beta[0-6]=\beta[9-15]$ and $\beta[7]=\beta[8]=0$. Then the key-recovery attack on the last 7-round KASUMI is proceeded with Partial-sum technique as follows:
\begin{figure}
  \centering
  \includegraphics[width=10cm]{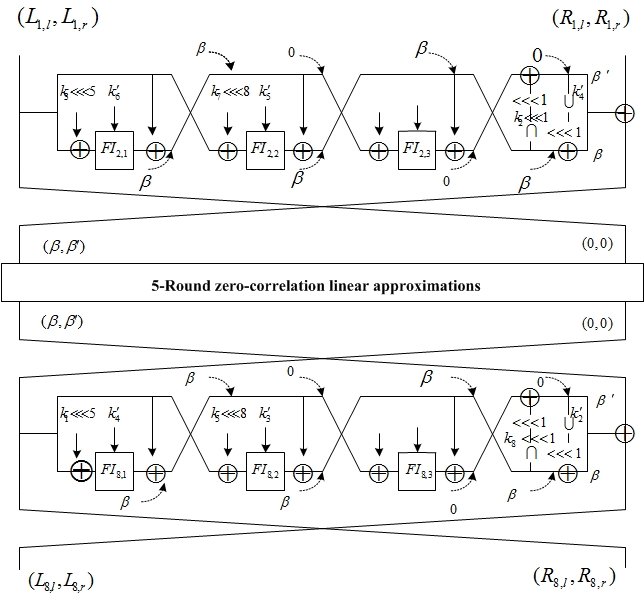}
  \caption{ Multidimensional Zero-correlation attack on KASUMI reduced to rounds 2-8}
\end{figure}

1. Collect $N$ plaintexts with corresponding ciphertexs.  Allocate a 8-bit counter $V_0[y_0]$ for each of $2^{85}$ possible values of $y_0=y_0^1\|y_0^2\|y_0^3\|y_0^4\|y_0^5\|y_0^6$ where $y_0^1 =L_{1,l}$, $y_0^2=L_{1,r}$,
$y_0^3 =R_{8,l}$, $y_0^4=R_{8,r}$,
$y_0^5 =(R_{1,l}\oplus L_{8,l}\oplus R_{8,r}\oplus L_{1,r})[0-6]\oplus (R_{1,l}\oplus L_{8,l}\oplus R_{8,r}\oplus L_{1,r})[9-15]$, $y_0^6=(L_{8,r}\oplus R_{1,l})[0-7]\|(R_{1,l}\oplus L_{8,l})[10-15]$, and set them zero. Calculate the
number of pairs of plaintext-ciphertext with given values $y_0$ and save it in $V_0[y_0]$. In
this step, around $2^{64}$ plaintext-ciphertext pairs are divided into $2^{85}$ different states.  So the assumption $V_0$ as a 8-bit counter is sufficient.

2. Guess the 46-bit $KO_{2,1}$, $KO_{8,1}$, $KI_{2,1}[0-6]$ and $KI_{8,1}[0-6]$. Allocate a counter $V_1[y_1]$ for each of $2^{53}$ possible values of $ y_1= y_1^1\|y_1^2\|y_1^3\|y_1^4$ where $y_1^1 =y_{0}^{2}$, $y_1^2 =y_{0}^{4}$, $y_1^3 =y_{0}^{6}$, and set them zero. For all $2^{32} $ possible values of $y_0^1$ and $y_0^3$, compute $y_1^4=y_0^5\oplus FI_{2,1}(y_0^1\oplus KO_{2,1}, KI_{2,1})[0-6]\oplus FI_{8,1}(y_0^3\oplus KO_{8,1}, KI_{8,1})[9-15]$ and update the value $V_1[y_1] = V_1[y_1] + V_0[y_0]$.

3. Guess the 7-bit $KL_{2,2}[7], KL_{2,2}[10-15]$ and then deduce $KL_{2,2}[0-6]$ from $KI_{8,1}[0-6]$. Allocate a counter $V_2[y_2]$ for each of $2^{39}$ possible values of $y_2=y_2^1\|y_2^2\|y_2^3$ where $y_2^1 = y_1^1$, $y_2^2 = y_1^2$ and set them zero. For all $2^{14} $ possible values of $y_1^3$, compute $y_2^3= y_1^4\oplus \Big(\big(\neg (KL_{2,2}[0-7]\| KL_{2,2}[10-15])\diamond y_1^3\big)\lll1\Big)[0-6]\oplus\Big(\big(\neg (KL_{2,2}[0-7]\| KL_{2,2}[10-15])\diamond y_1^3\big)\lll1\Big)[9-15]$ and update the value $V_2[y_2] = V_2[y_2] + V_1[y_1]$.

4. Guess the 16-bit $KO_{8,2}$ and  deduce $KI_{8,2}$ from $KO_{2,1}$. Allocate a counter $V_3[y_3]$ for each of $2^{23}$ possible values of
$y_3=y_3^1\|y_3^2$ where  $y_3^2 = y_2^1$ and set them zero. For all $2^{16} $ possible values of $y_2^2$, 
compute $y_3^1=y_2^3\oplus FI_{8,2}(y_2^2\oplus KO_{8,2}, KI_{8,2})[0-6]\oplus FI_{8,2}(y_2^2\oplus KO_{8,2}, KI_{8,2})[9-15]$ and update the value $V_3[y_3] = V_3[y_3] + V_2[y_2]$.

5. Guess the 16-bit $KO_{2,2}$ and  deduce $KI_{2,2}$ from $KO_{8,2}$. Allocate a counter $V_4[y_4]$ for each of $2^{7}$ possible values of
 $y_4$ and set them zero. For all $2^{16} $ possible values of $y_3^2$, compute 
 $y_3=y_3^1\oplus FI_{2,2}(y_3^2\oplus KO_{2,2}, KI_{2,2})[0-6]\oplus FI_{2,2}(y_3^2\oplus KO_{2,2}, KI_{2,2})[9-15]$
 and update the value $V_4[y_4] = V_4[y_4] + V_3[y_3]$.

6. Allocate 64-bit counters $V [z]$ for 7-bit z and initialize them to zero, where $z$
is the concatenation of evaluations of 7 basis zero-correlation masks. Compute $z$ from $y_4$ with 7 basis zero-correlation masks,
save it in $V[z]$, that is  $V [z]+ = V_4[y_4]$. Compute the statistic $T$ according to Equation (1). If $T< \tau$ , then the guessed key value is a right key candidate.

7. As there are 29 master key bits that we have not guessed, we do exhaustive search for all keys conforming to this possible key candidate.

In this attack, we set the type-I error probability $\beta_0 =2^{-2.7}$ and the type-II error probability $\beta_1 =2^{-5}$. We have
$z_{1-\beta_0}\approx 1$, $z_{1-\beta_1}\approx 2.4$, $n = 64$, $ l=2^{7}$. The date complex $N$ is about $2^{62.6}$ and the decision threshold $\tau \approx 2^{6.47}$.

There are $2^{85}$ master key value guessed during the encryption and decryption phase, and $2^{85}\cdot 2^{-5} = 2^{80}$ key candidates can survive in the wrong key filtration.  The complexity of the  Step 2, 3, 4, 6 and 7 is about $2^{62.6}\times 2^{46}=2^{108.6}$, $2^{46}\times 2^{7}\times 2^{53}=2^{106}$, $2^{46}\times 2^{7}\times 2^{16} \times 2^{39}=2^{108}$ and $2^{46}\times 2^{7}\times 2^{16} \times 2^{16} \times 2^{23}=2^{108}$ and $2^{80}\times 2^{29}=2^{109} $ 7-round KASUMI encryptions. the compute complexity is about $2^{110.5}$ 7-round KASUMI encryptions with $2^{62.1}$ known plaintexts and $2^{85}$ memory bytes for counters.

\section{\large\bf Conclusion }
In this paper, we evaluate the security of KASUMI with respect to the novel technique of the multidimensional zero-correlation cryptanalysis.
We investigate the properties of the linear masks propagate in  components (AND, OR functions)and then show some observations on the $FL$,$FO$ and $FI$ function. By selecting some special input/output masks, we refine the zero-correlation linear approximations and give first multidimensional zero-correlation attack on the 6-round KASUMI.  Moreover, under the weak keys conditions that the second keys of the $FL$ function in round 2 and round 8 have the same value at 1st to 8th and 11th to 16th bit-positions,  the paper expanded the attack to 7 rounds(2-8). The two attacks need $2^{85}$ encryptions with $2^{62.8}$ chosen plaintexts, $2^{54}$ memory bytes and $2^{110.5}$ encryptions with $2^{62.1}$ known plaintexts, $2^{85}$ memory bytes, respectively.

\vspace{0.1in}

\leftline{\bf References} \bigbreak
\def\REF#1{\par\hangindent\parindent\indent\llap{#1\enspace}\ignorespaces}
\footnotesize
\small
\REF{[1]} Bogdanov, A.,  Rijmen, V.: Linear Hulls with Correlation Zero and Linear Cryptanalysis of Block Ciphers. Designs, Codes and Cryptography, Springer, US, 2012, pp.1-15.
\REF{[2]} Bogdanov, A., Wang, M.: Zero Correlation Linear Cryptanalysis with
Reduced Data Complexity, in: A. Canteaut (Ed.), FSE 2012, in: Lect.Notes Com put. Sci., vol. 7549, Springer, Heidelberg, 2012, pp. 29-48.
\REF{[3]}Bogdanov, A., Leander, G., Nyberg, K., Wang, M. : Integral and multidimensional linear distinguishers with correlation zero, in: X. Wang,K. Sako (Eds.), AsiaCrypt 2012, in: Lect. Notes Comput. Sci., vol. 7658,
Springer, Heidelberg, 2012, pp. 24-262.
\REF{[4]} Bogdanov, A., Geng, H., Wang, M.,  Wen, L., Collard, B.: Zero-correlation
linear cryptanalysis with FFT and improved attacks on ISO standards
Camellia and CLEFIA, in: T. Lange, K. Lauter, P. Lisonek (Eds.), SAC¡¯13,
in: Lect. Notes Comput. Sci., Springer-Verlag, 2013, in press.
\REF{[5]}Biham, E., Dunkelman, O., Keller, N.: A Related-Key Rectangle Attack on the
Full KASUMI. In: Roy, B. (ed.) ASIACRYPT 2005. LNCS, vol. 3788, pp. 443-461.
\REF{[6]} Blunden, M., Escott, A.: Related Key Attacks on Reduced Round KASUMI. In:
Matsui, M. (ed.) FSE 2001. LNCS, vol. 2355, pp. 277-285.
\REF{[7]} Dunkelman, O., Keller, N., Shamir, A.: A Practical-Time Related-Key Attack on
the KASUMI Cryptosystem Used in GSM and 3G Telephony. In: Rabin, T. (ed.)
CRYPTO 2010. LNCS, vol. 6223, pp. 393-410.
\REF{[8]} Ferguson, N., Kelsey, J., Lucks, S., Schneier, B., Stay, M., Wagner, D., Whiting, D.: Improved cryptanalysis of Rijndael.  FSE 2000. LNCS, vol. 1978, pp. 213-230.
\REF{[9]} Jia, K., Li,L., Rechberger, C., Chen, C., Wang, X.: Improved Cryptanalysis of the Block Cipher KASUMI. In:
Knudsen, L.R., Wu, H. (eds.) SAC 2012. LNCS, vol. 7707, pp. 222-233. Springer, Heidelberg (2012)
\REF{[10]} K\" uhn, U.: Cryptanalysis of Reduced-Round MISTY. In: Pfitzmann, B. (ed.) EUROCRYPT 2001. LNCS, vol. 2045, pp. 325-339.
\REF{[11]} Matsui, M.: New Block Encryption Algorithm MISTY. In: Biham, E. (ed.) FSE 1997. LNCS, vol. 1267, pp. 54-68.
\REF{[12]}Sugio, N., Aono, H., Hongo, S., Kaneko, T.: A Study on Integral-Interpolation
Attack of MISTY1 and KASUMI. In: Computer Security Symposium 2006, pp.173-178.
\REF{[13]} Sugio, N., Aono, H., Hongo, S., Kaneko, T.: A Study on Higher Order Differential
Attack of KASUMI. IEICE Transactions 90-A(1), pp:14-21 (2007).
\REF{[14]} Sugio, N., Tanaka, H., Kaneko, T.: A Study on Higher Order Differential Attack
of KASUMI. In: 2002 International Symposium on Information Theory and its
Applications (2002).
\REF{[15]}3rd Generation Partnership Project, Technical Specification Group Services and
System Aspects, 3G Security, Specification of the 3GPP Confidentiality and Integrity Algorithms; Document 2: KASUMI Specification, V3.1.1 (2001).
\REF{[16]} 3rd Generation Partnership Project, Technical Specification Group Services and system Aspects, 3G Security,Specification of the A5/3 Encryption Algorithms for GSM and ECSD, and the GEA3 Encryption Algorithm for GPRS; Document 1: A5/3 and GEA3 Specifications, V6.2.0 (2003)
\REF{[17]}Wen, L., Wang, M., Bogdanov, A., Chen,H.: Multidimensional Zero-Correlation Attacks on Lightweight Block Cipher HIGHT: Improved Cryptanalysis of an ISO Standard. Information Processing Letters 114(6), pp. 322-330.
\REF{[18]}Wen, L.,Wang, M., Bogdanov, A.: Multidimensional Zero-Correlation Linear Cryptanalysis of E2. Africacrypt'14, Lecture Notes in Computer Science (LNCS), Springer-Verlag, 2014, to appear.

\end{document}